\documentclass[aps,prd,preprint,superscriptaddress,amsmath,amssymb,showpacs]{revtex4-1}
\usepackage{todonotes}
\usepackage{dcolumn}
\usepackage{graphicx}
\usepackage{float}
\usepackage{physics}
\usepackage[colorlinks=true,allcolors=blue]{hyperref}
\usepackage{epstopdf}
\usepackage{changes}
\usepackage{xcolor}

\begin{document}

\title{Potential energy of heavy quarkonium in flavor-dependent systems from a holographic model}

\author{Xi Guo}
\affiliation{School of Nuclear Science and Technology, University of South China Hengyang, China Number 28, West Changsheng Road, Hengyang City, Hunan Province, China.}

\author{Xun Chen}
\email{chenxunhep@qq.com}
\affiliation{School of Nuclear Science and Technology, University of South China Hengyang, China Number 28, West Changsheng Road, Hengyang City, Hunan Province, China.}

\author{Dong Xiang}
\email{xiangdong@usc.edu.cn}
\affiliation{School of Nuclear Science and Technology, University of South China Hengyang, China Number 28, West Changsheng Road, Hengyang City, Hunan Province, China.}

\author{Miguel Angel Martin Contreras}
\email{miguelangel.martin@usc.edu.cn}
\affiliation{School of Nuclear Science and Technology, University of South China Hengyang, China Number 28, West Changsheng Road, Hengyang City, Hunan Province, China.}

\author{Xiao-Hua Li}
\email{lixiaohuaphysics@126.com}
\affiliation{School of Nuclear Science and Technology, University of South China Hengyang, China Number 28, West Changsheng Road, Hengyang City, Hunan Province, China.}
%\date{\today}

\begin{abstract}
Within the framework of the Einstein-Maxwell-Dilaton model, which incorporates information on the equation of state and baryon number susceptibility from lattice results, we have conducted a comprehensive analysis of the potential energy, running coupling, and dissociation time for heavy quark-antiquark pairs using gauge/gravity duality. This study encompasses various systems, including pure gluon systems, 2 flavor systems, 2+1 flavor systems, and 2+1+1 flavor systems under finite temperature and chemical potential. The results reveal that the linear component of the potential energy diminishes as the flavor increases. It is also found that our results are extremely close to the recent lattice results for 2+1 flavors at finite temperature. Moreover, we have thoroughly investigated the dissociation distance and effective running coupling constant of quark-antiquark pairs to gain a comprehensive understanding of their behavior across various flavors. Finally, we have examined real-time dynamics of quark dissociation. The findings indicate that the dissociation time of quark-antiquark pairs is dependent on temperature, chemical potential, and flavor of the systems.
\end{abstract}

\maketitle

\maketitle
\section{Introduction}\label{sec-int}
The investigation of heavy quarks serves as a highly sensitive method in experimental research on quark-gluon plasma (QGP) and its characteristics. These heavy quarks act as significant probes for examining QCD matter \cite{Bulava:2019iut,Li:2023ciq,Yang:2015aia,Zhou:2020ssi}. The dissociation of heavy quark-antiquark pairs is widely regarded as a hallmark of deconfinement, consequently making the quark-antiquark potential a topic of great interest in the realm of QCD. The holographic potential for quark-antiquark pairs was initially documented in Ref.~\cite{Maldacena:1998im}. The interquark potential plays a crucial role in determining the formation of bound states of baryons, and studying it contributes to a better understanding of baryon structure and the dynamic mechanisms of QCD \cite{Andreev:2006ct,Andreev:2006eh,Andreev:2006nw,He:2010bx,Colangelo:2010pe,DeWolfe:2010he,Li:2011hp,Fadafan:2011gm,Fadafan:2012qy,Cai:2012xh,Li:2012ay,Fang:2015ytf,Yang:2015aia,Zhang:2015faa,Ewerz:2016zsx,Chen:2017lsf,Arefeva:2018hyo,Chen:2018vty,Bohra:2019ebj,Chen:2019rez,Zhou:2020ssi,Zhou:2021sdy,Chen:2020ath,Chen:2021gop}. Therefore, the objective of this study is to investigate the correlation properties of quarkonium pairs by utilizing the quark potential.

Relativistic heavy-ion collision experiments generate extreme conditions such as high temperatures and densities. Nevertheless, the study of lattice QCD at finite chemical potential continues to pose challenges due to the presence of the fermion sign problem \cite{Fodor:2001au,Muroya:2003qs}. Various approaches have been suggested to overcome this hurdle \cite{Lang:1982tj,Hoek:1987uy,Michael:1990az,Takahashi:2002bw,Aoki:2005vt,Ratti:2005jh,Bicudo:2007xp,Luscher:2010iy,Hasenfratz:1983ba,Jiang:2022zbt,Yu:2023hzl}. Although several methods have been developed so far, the study of strongly coupled Yang-Mills theories, such as QCD, continues to pose a challenge. A successful string description of realistic QCD has not yet been achieved. Numerous "top-down" approaches are currently being pursued to derive a realistic description of holographic QCD from string theory \cite{He:2007juu,Burrington:2004id,Sakai:2005yt,Sakai:2004cn,Erdmenger:2020lvq}. Conversely, the "bottom-up" approach uses experimental data and lattice results to build a holographic model \cite{Braga:2017fsb,Braga:2018fyc,Ferreira:2019nkz,He:2010ye,Jokela:2024xgz,Arefeva:2023fky,Rougemont:2023gfz}. Introducing a black hole in five-dimensional space to describe the boundary theory at finite temperature, as well as exploring more general backgrounds, are among the aspects being considered \cite{Rey:1998bq,Brandhuber:1998bs,Noronha:2010hb,Fadafan:2012qy,Zhang:2015faa,Zhang:2016jns,Chen:2021bkc,Guo:2023zjx}.

Additionally, we have also investigated the real-time dynamics of quark pair dissociation in QCD media \cite{Iatrakis:2015sua,Jena:2022nzw}. From a holographic perspective, the background geometry is described by a black hole spacetime at a specific temperature. The quark-antiquark pairs are positioned on the boundary with a separation distance of $L$, connected by a string \cite{Matsui:1986dk,Song:2007gm,Escobedo:2013tca}. Initially, the string hangs on the boundary and is subsequently attracted towards the black hole horizon by the gravitational forces arising from the background metric. As the string approaches the horizon, the system reaches an equilibrium state, and the string undergoes a dissociation process. The dissociation time of quark-antiquark pairs serves as a crucial timescale in the study of quarkonium suppression physics, providing valuable insights into the underlying mechanisms of QGP formation.

Based on the holographic QCD model described in Refs. ~\cite{Chen:2024ckb,Chen:2024mmd}, the primary objective of this study is to explore the impact of temperature and chemical potential on the potential of quarkonium pairs in various systems. These systems encompass pure gluon systems, 2 flavor systems, 2+1 flavor systems, and 2+1+1 flavor systems. Furthermore, this research aims to analyze the dissociation time of quarkonium pairs when temperature and chemical potential are introduced.

The remaining sections of this paper are organized as follows: Sec. \ref{sec2} provides a short review of the machine-learning holographic model that incorporates the information of QCD phase transition. In Sec. \ref{sec3}, we have conducted computational calculations and in-depth analyses of the potential variations of quark-antiquark pairs under varying temperature and chemical potential conditions. Our investigation covers a range of systems, including pure gluon systems, 2 flavor system, 2+1 flavor system, and 2+1+1 flavor system. Additionally, we have computed the dissociation distance and running coupling constant of quark-antiquark pairs in each scenario to provide comprehensive insights into their behavior. The interaction of quark-antiquark pair, namely effective running coupling, is studied for various systems in Sec. \ref{sec4}.  Section. \ref{sec5} focuses on the computation of the dissociation time of quarkonium pairs under various temperature and chemical potential scenarios. Finally, in Sec. \ref{sec6}, we present a succinct summary of the research findings and provide our concluding remarks.

\section{Setup}
\label{sec2}
The Einstein-Maxwell-dilaton model(EMD) will be reviewed in the section. In the string frame, the action of EMD model is expressed by \cite{He:2013qq,Yang:2014bqa,Yang:2015aia,Dudal:2017max,Dudal:2018ztm,Chen:2018vty,Chen:2020ath,Zhou:2020ssi,Chen:2019rez,Chen:2024ckb,Chen:2024mmd}
\begin{equation}
S_b=\frac{1}{16 \pi G_5} \int d^5 x \sqrt{-g^s} e^{-2 \phi_s}\left[R_s-\frac{f_s\left(\phi_s\right)}{4} F^2+4 \partial_\mu \phi_s \partial^\mu \phi_s-V_s\left(\phi_s\right)\right],
\label{Eq:actionsb}
\end{equation}
where $f(\phi)$ is the gauge kinetic function coupled with the Maxwell field $A_\mu$, $V\left(\phi\right)$ is the potential of the dilaton field, and $G_5$ is the Newton constant in five dimensions. The explicit forms of the gauge kinetic function $f\left(\phi\right)$ and the dilaton potential $V\left(\phi\right)$ can be solved consistently from the equations of motion(EoMs).
We transform the action from string frame to Einstein frame with the following transformations.
\begin{equation}
\phi_s=\sqrt{\frac{3}{8}} \phi, \quad g_{\mu \nu}^s=g_{\mu \nu} e^{\sqrt{\frac{2}{3}} \phi}, \quad f_s\left(\phi_s\right)=f(\phi) e^{\sqrt{\frac{2}{3}} \phi}, \quad V_s\left(\phi_s\right)=e^{-\sqrt{\frac{2}{3}} \phi} V(\phi).
\end{equation}
The action in Einstein frame becomes
\begin{equation}
\begin{aligned}
S_b & =\frac{1}{16 \pi G_5} \int d^5 x \sqrt{-g}\left[R-\frac{f(\phi)}{4} F^2-\frac{1}{2} \partial_\mu \phi \partial^\mu \phi-V(\phi)\right]. \\
\end{aligned}
\end{equation}
Then, we give the following ansatz of metric
\begin{eqnarray}
d s^2=\frac{R_{AdS}^2 e^{2 A(z)}}{z^2}\left[-g(z) d t^2+\frac{d z^2}{g(z)}+d \vec{x}^2\right],
\end{eqnarray}
where $z$ denotes the fifth-dimensional holographic coordinate, and the radial parameter $R_{AdS}$ of $\rm{AdS_5}$ space is fixed at $R_{AdS} = 1$. Using the above ansatz of the metric, the EoMs and constraints for the background fields can be obtained as
\begin{equation}
\begin{gathered}
\phi^{\prime \prime}+\phi^{\prime}\left(-\frac{3}{z}+\frac{g^{\prime}}{g}+3 A^{\prime}\right)-\frac{ e^{2 A}}{z^2 g} \frac{\partial V}{\partial \phi}+\frac{z^2 e^{-2 A} A_t^{\prime 2}}{2 g} \frac{\partial f}{\partial \phi}=0,
\end{gathered}
\end{equation}
\begin{equation}
A_t^{\prime \prime}+A_t^{\prime}\left(-\frac{1}{z}+\frac{f^{\prime}}{f}+A^{\prime}\right)=0,
\end{equation}
\begin{equation}
g^{\prime \prime}+g^{\prime}\left(-\frac{3}{z}+3 A^{\prime}\right)-e^{-2 A} A_t^{\prime 2} z^2 f=0,
\end{equation}
\begin{equation}
\begin{aligned}
A^{\prime \prime} & +\frac{g^{\prime \prime}}{6 g}+A^{\prime}\left(-\frac{6}{z}+\frac{3 g^{\prime}}{2 g}\right)-\frac{1}{z}\left(-\frac{4}{z}+\frac{3 g^{\prime}}{2 g}\right)+3 A^{\prime 2} +\frac{ e^{2 A} V}{3 z^2 g}=0,
\end{aligned}
\end{equation}
\begin{equation}
A^{\prime \prime}-A^{\prime}\left(-\frac{2}{z}+A^{\prime}\right)+\frac{\phi^{\prime 2}}{6}=0,
\end{equation}
where only four of the above five equations are independent.
The boundary conditions near the horizon are
\begin{equation}
A_t\left(z_h\right)=g\left(z_h\right)=0.
\end{equation}
Near the boundary $z \rightarrow z_h$, we require the metric in the string frame to be asymptotic to $\rm AdS_5$. The boundary conditions are
\begin{equation}
A(0)=-\sqrt{\frac{1}{6}} \phi(0), \quad g(0)=1, \quad A_t(0)=\mu+\rho^{\prime} z^2+\cdots.
\end{equation}
$\mu$ can be regarded as baryon chemical potential and $\rho^{\prime}$ is proportional to the baryon number density. $\mu$ is related to the quark-number chemical potential $\mu = 3\mu_q$.
Then, we can get
\begin{equation}
\begin{aligned}
g(z)&=1-\frac{1}{\int_0^{z h} d x x^3 e^{-3 A(x)}}\left[\int_0^z d x x^3 e^{-3 A(x)}+\frac{2 c \mu^2 e^k}{\left(1-e^{-c z_h^2}\right)^2} \operatorname{det} \mathcal{G}\right],\\
\phi^{\prime}(z) & =\sqrt{6\left(A^{\prime 2}-A^{\prime \prime}-2 A^{\prime} / z\right)}, \\
A_t(z) & =\mu \frac{e^{-c z^2}-e^{-c z_h^2}}{1-e^{-c z_h^2}}, \\
V(z) & =-3 z^2 g e^{-2 A}\left[A^{\prime \prime}+A^{\prime}\left(3 A^{\prime}-\frac{6}{z}+\frac{3 g^{\prime}}{2 g}\right)-\frac{1}{z}\left(-\frac{4}{z}+\frac{3 g^{\prime}}{2 g}\right)+\frac{g^{\prime \prime}}{6 g}\right],
\end{aligned}
\end{equation}
where
\begin{equation}
\operatorname{det} \mathcal{G}=\left|\begin{array}{ll}
\int_0^{z_h} d y y^3 e^{-3 A(y)} & \int_0^{z_h} d y y^3 e^{-3 A(y)-c y^2} \\
\int_{z_h}^z d y y^3 e^{-3 A(y)} & \int_{z_h}^z d y y^3 e^{-3 A(y)-c y^2}
\end{array}\right|.
\end{equation}
The Hawking temperature is given by,
\begin{equation}
\begin{aligned}
T & =\frac{z_h^3 e^{-3 A\left(z_h\right)}}{4 \pi \int_0^{z_h} d y y^3 e^{-3 A(y)}}\Big[1+ \\
&\frac{2 c \mu^2 e^k\left(e^{-c z_h^2} \int_0^{z_h} d y y^3 e^{-3 A(y)}-\int_0^{z_h} d y y^3 e^{-3 A(y)} e^{-c y^2}\right)}{(1-e^{-c z_h^2})^2} \Big],
\end{aligned}
\end{equation}
To obtain an analytical solution, we employ
\begin{eqnarray}
A(z)=d \ln \left(a z^2+1\right)+d \ln \left(b z^4+1\right) \text {, }
\end{eqnarray}
and the gauge kinetic function $f(z)$ is taken as
\begin{equation}\label{fff}
f(z)=e^{c z^2-A(z)+k}.
\end{equation}
Then, we can calculate the potential of the quark-antiquark pair in the string frame with $A_s(z)=A(z)+\sqrt{\frac{1}{6}} \phi(z)$. In the string frame, we can use the standard process to obtain the separation distance and potential of quark-antiquark pairs\cite{Rey:1998bq,Li:2011hp,Colangelo:2010pe,Chen:2017lsf,Guo:2023zjx}.  The string world-sheet action is defined by the Nambu-Goto action and takes the following form
\begin{equation}
S_{N G}=-\frac{1}{2 \pi \alpha^{\prime}} \int \mathrm{d}^{2} \xi \sqrt{-\operatorname{det} g_{a b}}.
\end{equation}
Here, $g_{a b}$ is the induced metric defined as
\begin{equation}g_{a b}=g^s_{M N} \partial_{a} X^{M} \partial_{b} X^{N}, \quad a, b=0,1,
\end{equation}
and $\alpha'$ is related to the string tension and is set to 1. Here, $X^{M}$ and $g^s_{M N}$ are the coordinates and the string-frame metric, respectively.
To calculate the quark-antiquark potential, we consider the string ends at a static quark-antiquark pair locating at $x_1=-L/2$ and $x_1=L/2$. A simplest parametrization of the string world-sheet parameters is $\xi^{0}=t, \xi^{1}=x_1$. Under this condition, the effective Nambu-Goto action can be written as
\begin{equation}
S_{N G}=-\frac{1}{2 \pi  T} \int_{-L / 2}^{L / 2} \mathrm{d} x_1 \sqrt{k_{1}(z) \frac{\mathrm{d} z^{2}}{\mathrm{d} x_1^{2}}+k_{2}(z)},
\end{equation}
where
\begin{eqnarray}
\begin{aligned}
 k_{1}&=\frac{e^{4A_{s}}}{z^{4}},  \\
 k_{2}&=\frac{e^{4A_{s}}}{z^{4}} g(z).\\
\end{aligned}
\end{eqnarray}
According to the study in \cite{Witten:1998zw,Rey:1998bq,Brandhuber:1998er,Gross:1998gk,Andreev:2006nw,Giataganas:2011nz}, the expectation value of the Wigner-Wilson loop is then related to the on-shell string action by

\begin{equation}
\langle W(\mathcal{C})\rangle=\int D X e^{-S_{N G}} \simeq e^{-S_{on-shell}},
\end{equation}
where $\mathcal{C}$ denotes a closed loop in spacetime. The definition of the heavy-quark potential is \cite{Maldacena:1998im,Rey:1998ik,Ewerz:2016zsx}

\begin{equation}
\langle W(\mathcal{C})\rangle \sim e^{-V(r, T) / T},
\end{equation}
where $L$ is the separate distance of quarks. Thus, to get the potential, one has to solve the on-shell string world-sheet action. Following the standard procedure \cite{Li:2011hp,Andreev:2006nw,Colangelo:2010pe,Yang:2015aia,Chen:2017lsf,Chen:2020ath}, we can define an effecive `Hamitonian'
\begin{equation}
\mathcal{H} = z' \frac{\partial{\mathcal{L}}}{\partial{z'}} - \mathcal{L} = \frac{k_2(z)}{\sqrt{k_1(z) z'^2 + k_2(z)}},
\end{equation}
where $z'=\frac{\mathrm{d}z }{\mathrm{d}x_{1} }$. Solving  $z'$ forms the equation
\begin{equation} \frac{k_2(z)}{\sqrt{k_1(z) z'^2 + k_2(z)}} = \frac{k_2(z_0)}{\sqrt{k_2(z_0)}}. \end{equation}
Here $z_0$ denotes the position of the vertex where the quark-antiquark string joins, with values ranging from $z_0 =0$ to $z_0 =z_h$. Then, we can obtain the interquark distance and renormalized potential of heavy-quark-antiquark pair as
\begin{equation}
L=\int_{-\frac{L}{2}}^{\frac{L}{2}} d x=2 \int_{0}^{z_{0}} d z \frac{1}{z^{\prime}}= 2 \int_{0}^{z_{0}}\left[\frac{k_{2}(z)}{k_{1}(z)}\left(\frac{k_{2}(z)}{k_{2}\left(z_{0}\right)}-1\right)\right]^{-1 / 2} \mathrm{d} z,
\end{equation}
\begin{equation}
V =\frac{1}{\pi}\bigg(\int_{0}^{z_{0}} \mathrm{d} z(\sqrt{\frac{k_{2}(z) k_{1}(z)}{k_{2}(z)-k_{2}\left(z_{0}\right)}}-(\frac{1}{z^2}+\frac{2\sqrt{-6 a d}}{z}))-(\frac{1}{z_0}-2\sqrt{-6 a d} \ln(z_0)) \bigg).
\end{equation}
Here, we have regularized the potential by subtracting the divergent term at the UV. In this model, the parameters utilized are outlined in Table \ref{table:table1} from Refs.~\cite{Chen:2024ckb,Chen:2024mmd} to encode the thermodynamics of systems with different flavors. Next, a probe of heavy-quark pairs put in the various systems is investigated.

\begin{table}[h!]
  \begin{center}
    \begin{tabular}{|c|c|c|c|c|c|c|c|}
      \hline & a & b & c & d & k & $G_5$ & $T_c$ \\
      \hline $N_f$=0 & 0 & 0.072 & 0 & -0.584 & 0 & 1.326 & 0.265 \\
      \hline $N_f$=2 & 0.067 & 0.023 & -0.377 & -0.382 & 0 & 0.885 & 0.189 \\
      \hline $N_f$=2+1 & 0.204 & 0.013 & -0.264 & -0.173 & -0.824 & 0.400 & 0.128 \\
      \hline $N_f$=2+1+1 & 0.196 & 0.014 & -0.362 & -0.171 & -0.735 & 0.391 & 0.131 \\
      \hline
    \end{tabular}
    \caption{Parameters of pure gluon system, 2-flavor, 2+1-flavor, and 2+1+1-flavor system, respectively. $T_c$ is the predicted critical temperature at vanishing chemical potential. The unit of $T$ is GeV. }
    \label{table:table1}
  \end{center}
\end{table}

\section{Quarkonium potential and dissociation distance}
\label{sec3}
In this section, we mainly discuss the potential of quark-antiquark pairs and their dissociation distance at finite temperature and chemical potential in systems with various flavors. As the temperature and chemical potential increase, we will see that the quark-antiquark pairs are screened, resulting in their conversion into free quarks. Systems with different flavors will also exhibit different potential energies and dissociation distances.

\subsection{Different flavors at finite temperature}\label{dtem}
In previous studies, the real part of the in-medium potential is suggested to lie between the free energy and the internal energy, both of which exhibit a clear color screening effect as the temperature increases \cite{Chen:2024iil}. However, recent lattice QCD calculations with dynamic quarks suggest that there are no color screening effects for the real part of the heavy quark potential, even at high temperatures \cite{Bazavov:2023dci,Bala:2021fkm}. In the following, we will give a holographic calculation of heavy-quark potential for different flavors at finite temperature and compare with the recent lattice results from the Gaussian fits \cite{GaurangParkar:2022aoa}.

\begin{figure}
    \centering
    \includegraphics[width=15cm]{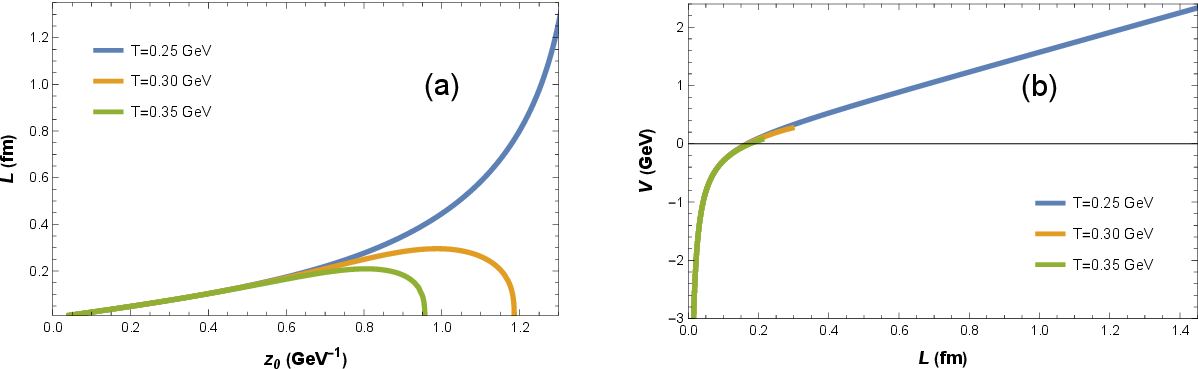}
    \caption{\label{fig1} (a) The dependence of interquark distance $L$ of a quark-antiquark pairs on $z_{0}$ at different temperatures for $N_{f}=0$, $\mu=0$. (b) The dependence of the potential energy $V$ of a quark-antiquark pairs on the interquark distance $L$ at different temperatures for $N_{f}=0$, $\mu=0$. }
\end{figure}

\begin{figure}
    \centering
    \includegraphics[width=15cm]{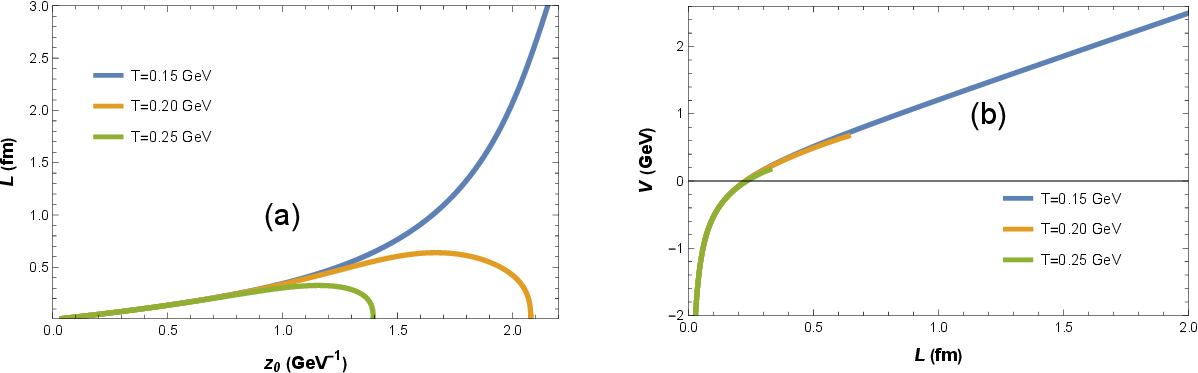}
    \caption{\label{fig2} (a) The dependence of interquark distance $L$ of a quark-antiquark pairs on $z_{0}$ at different temperatures for $N_{f}=2$, $\mu=0$. (b) The dependence of the potential energy $V$ of a quark-antiquark pairs on the interquark distance $L$ at different temperatures for $N_{f}=2$, $\mu=0$.  }
\end{figure}

\begin{figure}
    \centering
    \includegraphics[width=15cm]{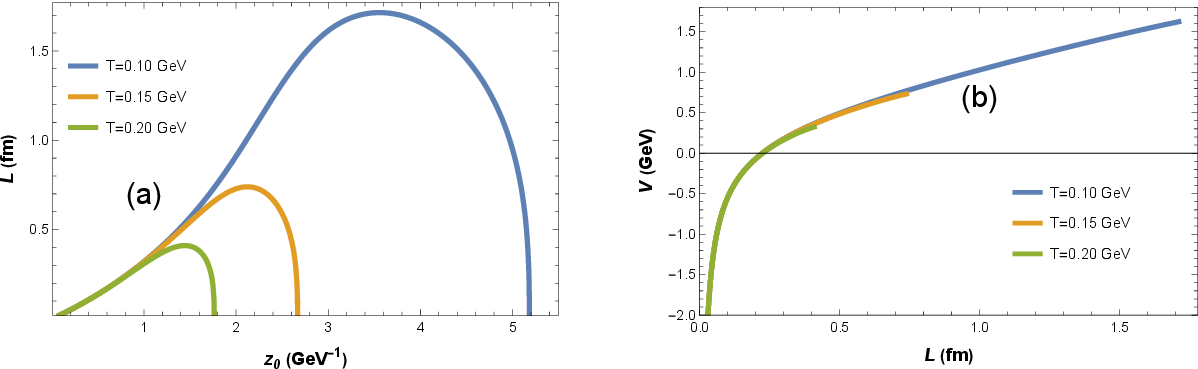}
    \caption{\label{fig3} (a) The dependence of interquark distance $L$ of a quark-antiquark pairs on $z_{0}$ at different temperatures for $N_{f}=2+1$, $\mu=0$. (b) The dependence of the potential energy $V$ of a quark-antiquark pairs on the interquark distance $L$ at different temperatures for $N_{f}=2+1$, $\mu=0$.  }
\end{figure}

\begin{figure}
    \centering
    \includegraphics[width=15cm]{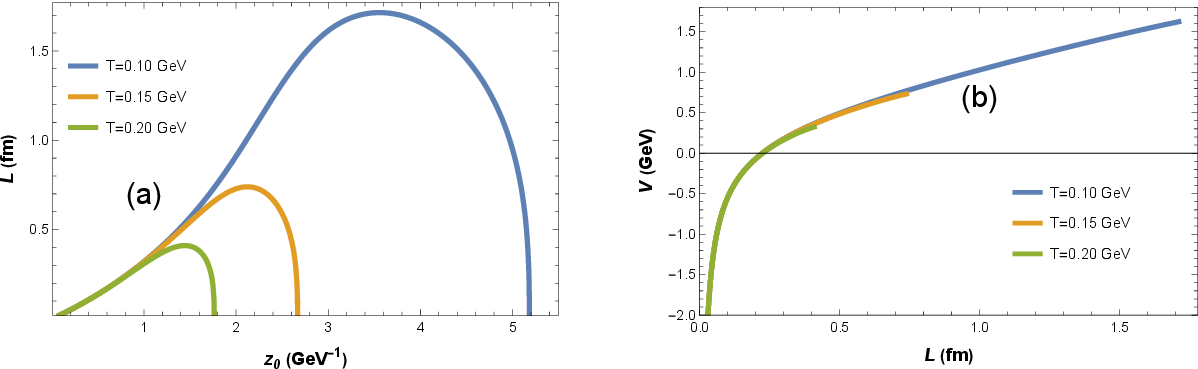}
    \caption{\label{fig4} (a) The dependence of interquark distance $L$ of quark-antiquark pairs on $z_{0}$ at different temperatures for $N_{f}=2+1+1$, $\mu=0$. (b) The dependence of the potential energy $V$ of quark-antiquark pairs on the interquark distance $L$ at different temperatures for $N_{f}=2+1+1$, $\mu=0$.   }
\end{figure}

\begin{figure}
    \centering
    \includegraphics[width=15cm]{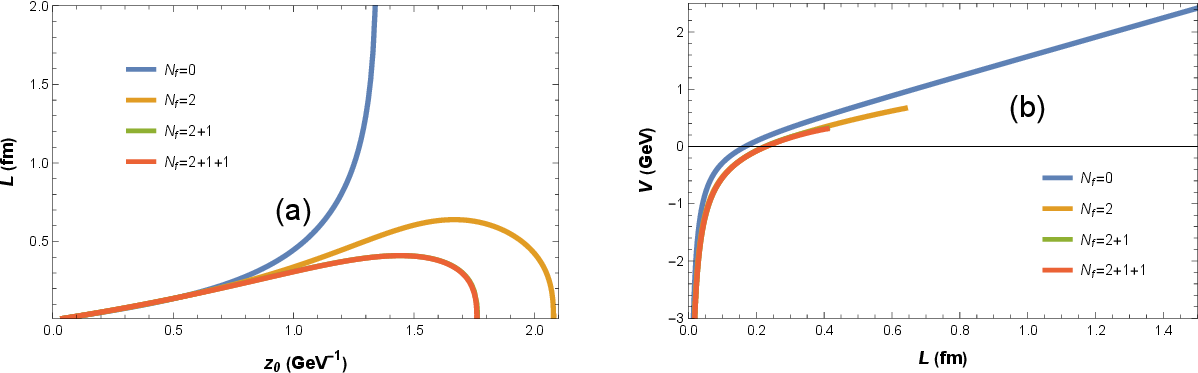}
    \caption{\label{fig5} (a) The dependence of interquark distance $L$ of quark-antiquark pairs on $z_{0}$ at different temperatures for $T=0.2$ GeV, $\mu=0$. (b) The dependence of the potential energy $V$ of quark-antiquark pairs on the interquark distance $L$ at different flavors for $T=0.2$ GeV, $\mu=0$.  }
\end{figure}

\begin{figure}
    \centering
    \includegraphics[width=8.5cm]{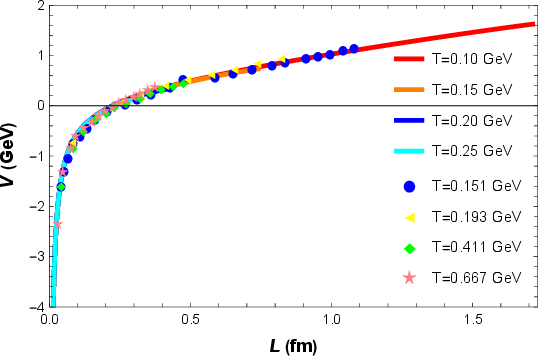}
    \caption{\label{fig6} Comparison of model data and lattice data \cite{GaurangParkar:2022aoa} for the quark-antiquark pairs potential at $N_{f}=2+1$, $\mu=0$. The different shapes of the points represent lattice data, while the lines depict model data.  }
\end{figure}

First, we calculate the properties of the pure gluon system, 2 flavor system, 2+1 flavor system, and 2+1+1 flavor system at finite temperatures shown in Figs.~\ref{fig1}-\ref{fig6}. Figure.~\ref{fig1} (a) shows the quark separation distance, $L$, increases with increasing $z_{0}$. Particularly, before reaching the critical temperature, the quark separation distance initially shows a slow increase with increasing $z_{0}$. However, as $z_{0}$ continues to increase, the separation distance suddenly expands. This behavior is reflected in the $V-L$ plot as an infinite linear increase in the potential energy of the quark-antiquark pair with increasing separation distance. Above the critical temperature, the quark separation distance increases with increasing $z_{0}$ until it reaches a maximum point ($L_{max}$). Beyond this point, the separation distance decreases with further increase in $z_{0}$. This suggests that the quark-antiquark pair enters the deconfined state, the string connecting them melts, and the quark-antiquark pair configuration no longer exists, transitioning into free quarks. For $N_{f}=0$ (Fig.\ref{fig1}(b)), the maximum dissociation distance for the quark pair at $T=0.3$ GeV is $0.295$ fm, while at $T=0.35$ GeV it is $0.208$ fm.

The potential energy consists of two components in Fig.~\ref{fig1} (b): the Coulomb potential at short distances and the linear potential at long distances. The temperature slightly affects the linear component of the potential energy. As the temperature increases, the linear part of the potential energy does not infinitely increase, but reaches a maximum point. This leads to a decrease in the linear component of the potential energy for the quark-antiquark pair. Specifically, in the $L-V$ plots for different flavors, it is evident that the temperature increase only affects the length of the linear component of the quark pair potential energy, while having little effect on the Coulomb potential component.

Figure.~\ref{fig2} (a) shows the quark separation distance, $L$, increases with increasing $z_{0}$ for the $N_{f}=2$. The qualitative behavior is the same as  before. However, the quark-antiquark pair will be screened under lower temperature. Compared with Fig.~\ref{fig1} (b), Fig.~\ref{fig2} (b) shows different behavior, indicating that the potential has a cutoff, which leads to a dissociation distance, under the same temperature $T = 0.25$ GeV.

Figures.~\ref{fig3} and~\ref{fig4} show the picture of separation distance and potential for 2+1 flavor and 2+1+1 flavor. It is found that the results are extremely close, as the presence of charm has only a slight influence on the thermodynamics of the medium within this temperature range.

To be more clear, we set the temperature to $T=0.2$ GeV and compared the separation distance and potential energy of quark pairs for different flavors, as shown in Fig.~\ref{fig5}. Consistent with our previous findings, these plots reveal that the separation distance between quark pairs initially increases and then decreases after reaching $L_{max}$ as $z_{0}$ increases. Interestingly, at $N_{f}=0$, there is no maximum separation distance. Instead, at larger values of $z_{0}$, there is a sharp increase in the separation distance of quark pairs. This phenomenon occurs because at $T=0.2$ GeV, the critical temperature of $N_{f}=0$ ($T_{c}=0.265$ GeV) has not been reached, causing the sudden increase in separation distance at larger $z_{0}$ values. Moreover, as the flavor increases, the dissociation distance $L_{max}$ between quarks decreases in Fig.~\ref{fig6} (a). Similarly, in the $V-L$ plots, the linear component of the potential energy for $N_{f}=0$ can extend to the infinity. In the 2-flavor system, 2+1-flavor system, and 2+1+1-flavor system, the potential energy decreases with an increasing flavor in Fig.~\ref{fig5} (b). This suggests that at the same temperature and chemical potential, large numbers of flavors are more prone to disrupt the string connecting the quark pair configuration, resulting in a less stable quark-antiquark pair.

Lattice QCD is a computational method used in gauge field theory, employing discretized spacetime grids for numerical simulations. It plays a critical role in theoretical physics and high-energy physics research, especially in studying quark-gluon plasma and nonperturbative effects of quantum chromodynamics. In this study, we compare the computational results of our model with lattice results of 2+1 flavor \cite{GaurangParkar:2022aoa}, as depicted in Fig.~\ref{fig6}. The data points represent lattice data, while the lines represent our model's computational results. The comparison reveals a striking similarity between the model's calculations and the lattice data without introducing extra parameters.

\subsection{Different flavors at finite chemical potential}\label{dmu}
\begin{figure}
    \centering
    \includegraphics[width=15cm]{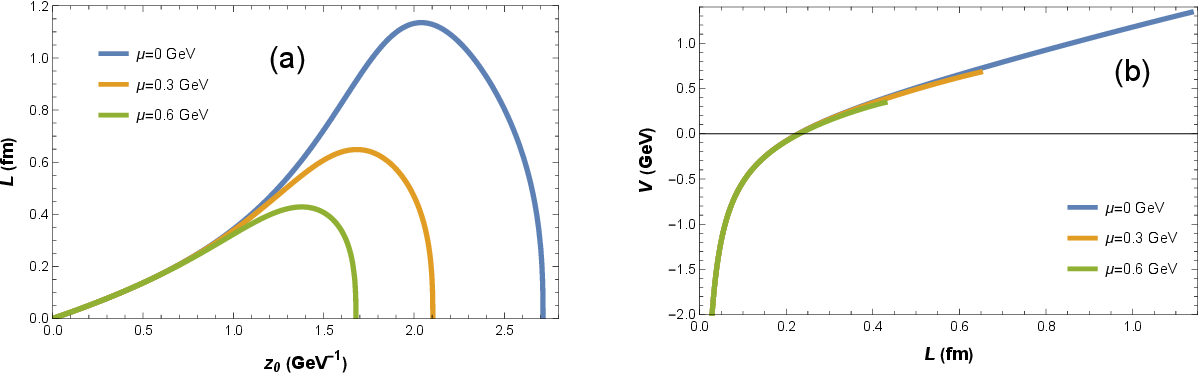}
    \caption{\label{fig7} (a) The dependence of interquark distance $L$ of quark-antiquark pairs on $z_{0}$ at different chemical potentials for $N_{f}=2$, $T=0.189$ GeV. (b) The dependence of the potential energy $V$ of quark-antiquark pairs on the interquark distance $L$ at different chemical potentials for $N_{f}=2$. }
\end{figure}

\begin{figure}
    \centering
    \includegraphics[width=15cm]{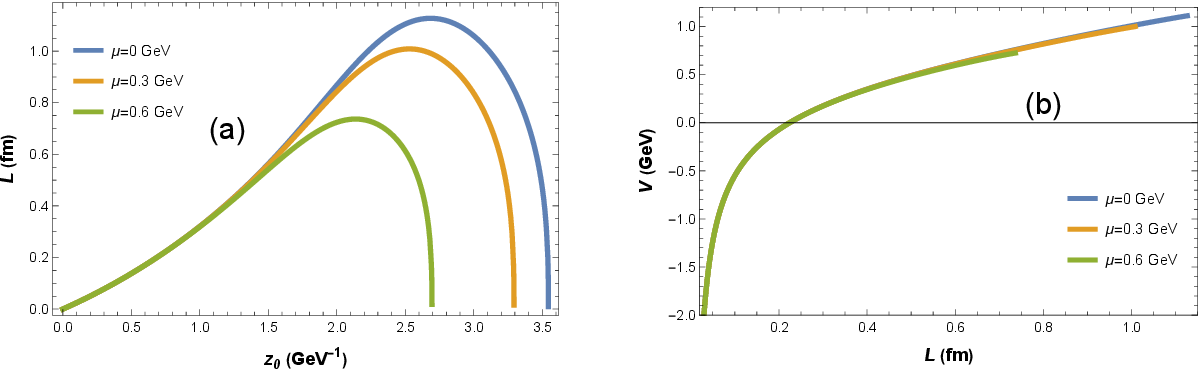}
    \caption{\label{fig8} (a) The dependence of interquark distance $L$ of quark-antiquark pairs on $z_{0}$ at different chemical potentials for $N_{f}=2+1$, $T=T_{c}=0.128$ GeV. (b) The dependence of the potential energy $V$ of quark-antiquark pairs on the interquark distance $L$ at different chemical potentials for $N_{f}=2+1$. }
\end{figure}

\begin{figure}
    \centering
    \includegraphics[width=15cm]{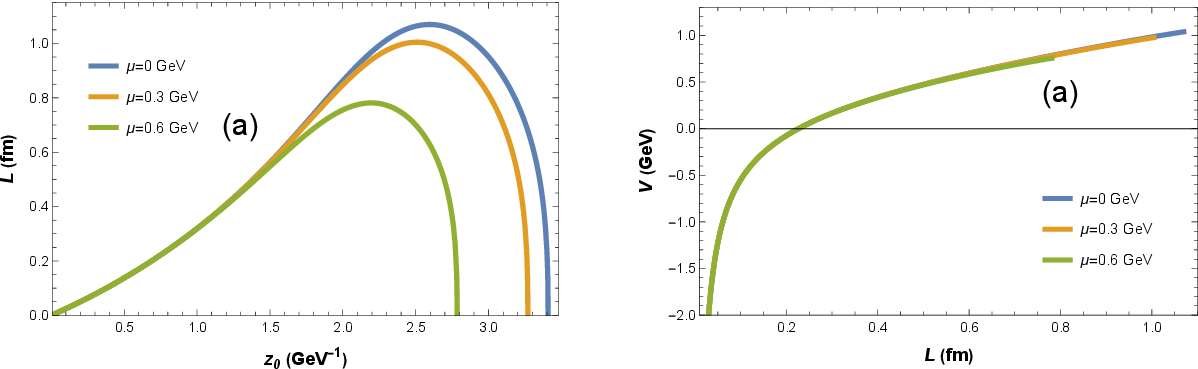}
    \caption{\label{fig9} (a) The dependence of interquark distance $L$ of quark-antiquark pairs on $z_{0}$ at different chemical potentials for $N_{f}=2+1+1$, $T=T_{c}=0.131$ GeV. (b) The dependence of the potential energy $V$ of quark-antiquark pairs on the interquark distance $L$ at different chemical potentials for $N_{f}=2+1+1$.}
\end{figure}

Similar to the previous section, we have computed the separation distance and potential energy of quark pairs for three different chemical potentials in 2-flavor systems, 2+1 flavor systems, and 2+1+1 flavor systems at their respective critical temperatures. The results are presented in Figs.~\ref{fig7}-~\ref{fig9}. In the $L-z_{0}$ plots for different flavors, we observe an initial increase in the separation distance $L$ with an increase in $z_{0}$. Similar to the temperature dependence, there exists a maximum value of $L_{max}$. Beyond this maximum separation distance, as $z_{0}$ continues to increase, the separation distance between quarks decreases. At this stage, the quark pairs are screened, and the string connecting them melts, resulting in free quarks as shown in Figs.~\ref{fig7} (a), \ref{fig8} (a), and \ref{fig9} (a). In the same flavor system, the increase in chemical potential causes $L_{max}$ to decrease, and the potential energy curve reaches a point where further increase in potential energy ceases. Nevertheless, when dealing with high chemical potentials, the potential energy curve only displays a slight deviation from the curve at low chemical potentials for the same separation distance, as depicted in Figs. \ref{fig7} (b) \ref{fig8} (b), and \ref{fig9} (b). This indicates that the presence of a chemical potential reduces the linear component of the quark pair's potential energy without affecting the Coulombic potential of the quark pair.

\begin{figure}
    \centering
    \includegraphics[width=15cm]{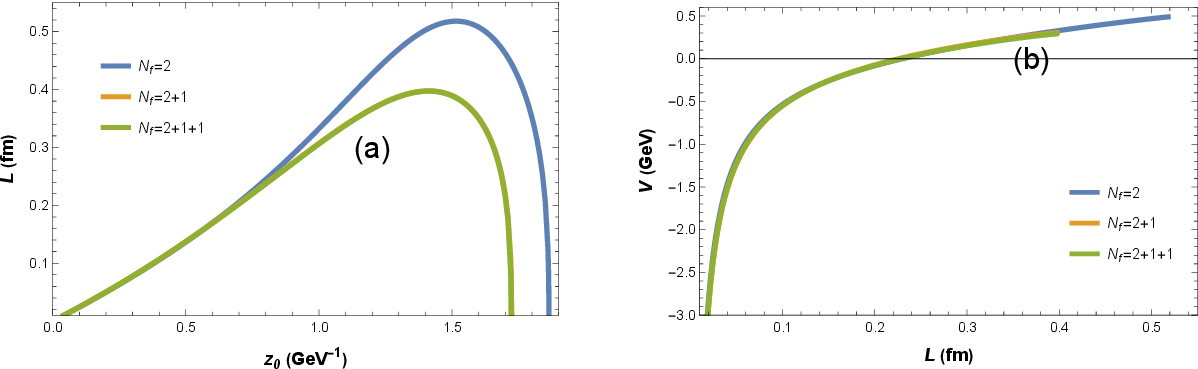}
    \caption{\label{fig10} (a) The dependence of interquark distance $L$ of quark-antiquark pairs on $z_{0}$ at different flavors for $\mu=0.3$ GeV, $T=0.2$ GeV. (b) The dependence of the potential energy $V$ of quark-antiquark pairs on the interquark distance $L$ at different flavors for $\mu=0.3$ GeV, $T=0.2$ GeV.}
\end{figure}

In Fig.~\ref{fig10}, we set the chemical potential to 0.3 GeV and the temperature to 0.2 GeV for each system. Subsequently, we calculate the plots illustrating the separation distance and potential energy of quark-antiquark pairs with different flavors. Under the same temperature and chemical potential, the maximum separation distances for quark pairs decrease with increasing flavor, and are 0.397, 0.398, and 0.518 fm, respectively. This indicates that as the number of flavors increases, the maximum separation distance of quark pairs also increases, suggesting their tendency to dissociate at shorter distances. In the potential energy plot, the endpoint distance for the potential decreases with the increase in the number of flavors, aligning with the findings from the quark separation distance plot. Besides, the results of the 2+1 and 2+1+1 flavors almost overlap, indicating that the effect of the 2+1 and 2+1+1 flavors on the heavy-quark potential is similar.

\subsection{The dissociation distance of quark-antiquark pairs }

In quark-antiquark pairs, a strong interaction force exists between the quark and antiquark. However, the presence of color charge in hadrons causes this interaction to gradually weaken over a specific distance until it becomes fully shielded. The dissociation distance defines the binding nature of the quark-antiquark pair. When the distance between the quark and antiquark is smaller than the dissociation distance, they are bound together through the strong interaction, resulting in the formation of stable hadrons like protons or mesons. Conversely, when the distance between the quark and antiquark exceeds the dissociation distance, their interaction is shielded, preventing them from being bound. This leads to the free quarks and antiquarks. Hence, the dissociation distance plays a crucial role in comprehending the binding of quark pairs and the mechanism of hadron formation. It aids in describing the properties of hadrons, decay processes, and elucidating particle production and deconfinement phenomena observed in high-energy physics experiments.

\begin{figure}
    \centering
    \includegraphics[width=8.5cm]{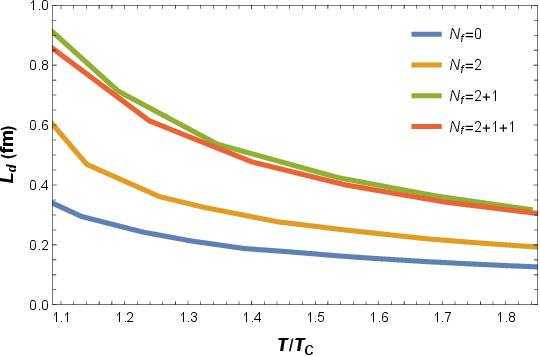}
    \caption{\label{fig11} The dissociation distance $L_{d}$ of quark-antiquark pairs is dependent on temperature $T$ ($T>T_c$) when $\mu=0$ GeV. }
\end{figure}

We have performed calculations and generated $L_{d}-T/T_{c}$ diagrams for different flavors at $\mu=0$, as depicted in Fig.~\ref{fig11}. The diagram demonstrates that with increasing temperature, the dissociation distance between quark pairs gradually diminishes. Notably, near the critical temperature, the dissociation distance experiences a rapid decline, whereas at temperatures far from the critical temperature, the decrease in the dissociation distance is more gradual across all systems. Furthermore, upon comparing the dissociation distance plots of the systems with different flavors, it becomes apparent that the different flavors of the system will significantly influence the dissociation distance of quarkonium.

\begin{figure}
    \centering
    \includegraphics[width=8.5cm]{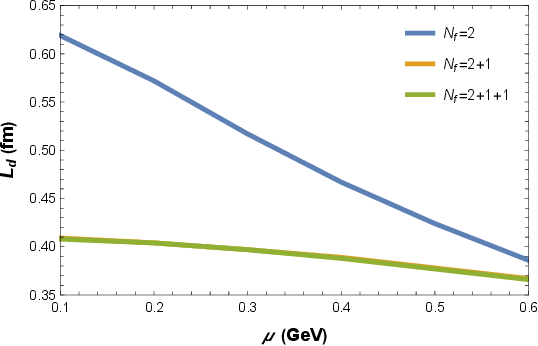}
    \caption{\label{fig12} The dissociation distance $L_{d}$ of quark-antiquark pairs is dependent on temperature $T$ when $T=0.2$ GeV. The unit of $L_{d}$ is fm and the unit of $\mu$ is GeV.}
\end{figure}

In Fig.~\ref{fig12}, we can observe the relationship between the dissociation distance of quark pairs and the chemical potential at the critical temperature in different flavor systems. As depicted, as the chemical potential rises, the dissociation distance of quark pairs gradually decreases, indicating a reduction in the maximum separation between quark pairs with increasing temperature. However, in contrast to the temperature dependence, as the chemical potential increases, the dissociation distance of quark pairs exhibits a slower rate of decrease compared to Fig.~\ref{fig11}. This suggests that temperature plays a more significant role than the chemical potential in influencing the dissociation distance. Additionally, we found that compared to the 2+1 flavor and 2+1+1 flavor systems, the chemical potential has a greater impact on the potential energy of heavy quarkonium in the 2 flavor system.


\section{Effective running coupling constant}
\label{sec4}
\begin{figure}
    \centering
    \includegraphics[width=8.5cm]{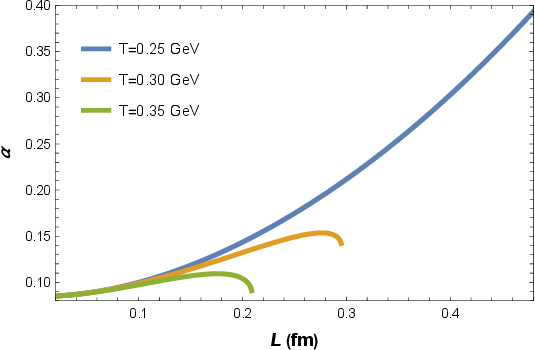}
    \caption{\label{fig13} The running coupling constant of the quark separation distance function, $L$, in $N_f=0$ at different temperatures.}
\end{figure}

\begin{figure}
    \centering
    \includegraphics[width=15cm]{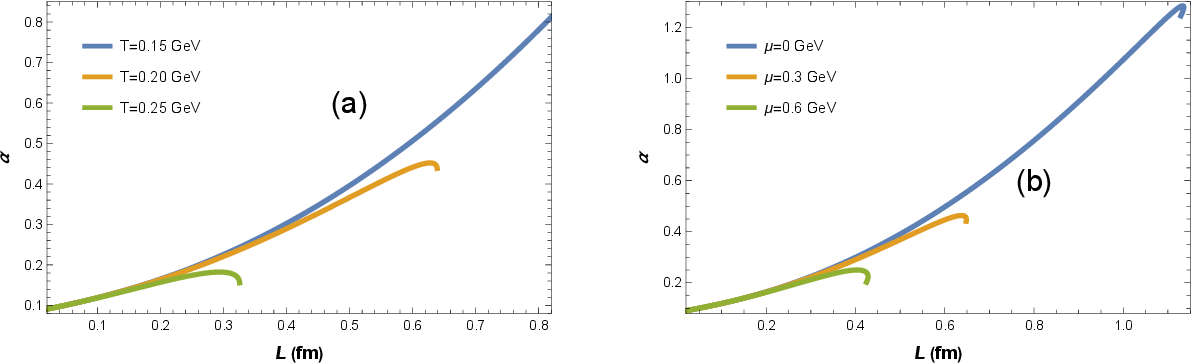}
    \caption{\label{fig14} (a) The running coupling constant of the quark separation distance function, $L$, in $N_f=2$ at different temperatures. (b) The running coupling constant of the quark separation distance function, $L$, in $N_f=2$ system at different chemical potential.}
\end{figure}

\begin{figure}
    \centering
    \includegraphics[width=15cm]{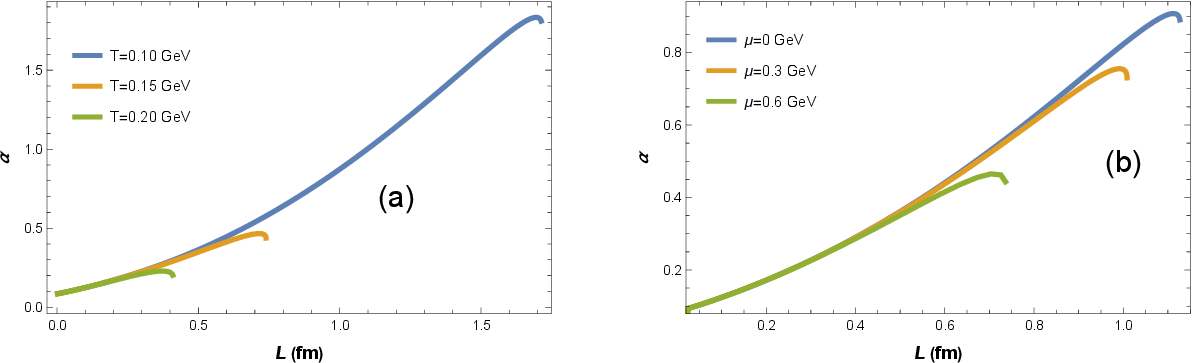}
    \caption{\label{fig15} (a) The running coupling constant of the quark separation distance function, $L$, in the $N_f=2+1$ at different temperatures. (b) The running coupling constant of the quark separation distance function, $L$, in the $N_f=2+1$ system at different chemical potentials.}
\end{figure}

\begin{figure}
    \centering
    \includegraphics[width=15cm]{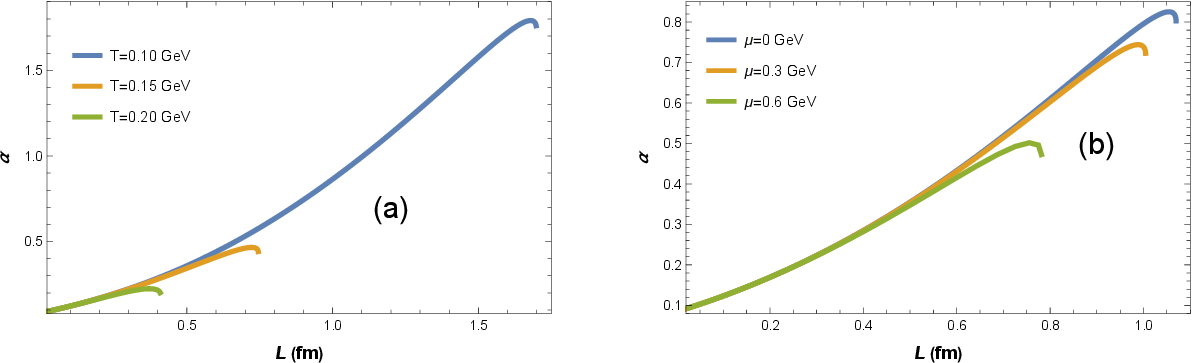}
    \caption{\label{fig16} (a) The running coupling constant of the quark separation distance function, $L$, in the $N_f=2+1+1$ at different temperatures. (b) The running coupling constant of the quark separation distance function, $L$, in the $N_f=2+1+1$ system at different chemical potentials.}
\end{figure}

The running coupling constant serves as a measure of the strength of interactions at different energy scales and is essential for making accurate theoretical predictions in various physical processes \cite{Deur:2016tte,Yu:2021yvw,DeSanctis:2023clb,Arefeva:2024vom}. At finite temperature and chemical potential, the QCD coupling indeed exhibits running behavior as a function of the quark separation distance $L$ in the holographic models  \cite{Ewerz:2012xgb,Ewerz:2015jno,Chen:2021gop,Zhou:2023qtr,Liu:2023hoq}. According to the definitions provided in lattice QCD \cite{Kaczmarek:2005ui,Kaczmarek:2004gv}, the effective running coupling constant is defined in the so-called qq-scheme as
\begin{eqnarray}
\alpha_{Q \bar{Q}}=\frac{3 L^2}{4} \frac{\mathrm{d} V_{Q \bar{Q}}}{\mathrm{~d} L}.
\label{eq:cou}
\end{eqnarray}
For Figs.~\ref{fig13}, \ref{fig14} (a), \ref{fig15} (a), and \ref{fig16} (a) of different systems at varying temperatures with $\mu=0$, we observed that before reaching the critical temperature of each system, the running coupling constant increases with the quark separation distance. However, the running coupling constant initially continues to rise with $L$, reaching a peak value, and then starts to decrease once the critical temperature is surpassed. The magnitude of the running coupling constant represents the strength of interactions for the heavy quarkonium. For strong interactions, the coupling constant is large, indicating a significant interaction force where interactions between particles cannot be ignored. Conversely, the small coupling constant indicates that interactions between quarks can be approximated as independent behavior.

Additionally, within the same system, the maximum value of the running coupling constant diminishes as the temperature rises. Hence, as the quark separation distance increases, the interaction force between quarks intensifies until reaching a maximum value, after which the interaction of heavy quarkonium becomes weak. Figures.~\ref{fig14} (b), \ref{fig15} (b), and \ref{fig16} (b) illustrate the $L-\alpha$ plots of different systems under various chemical potentials when $T=T_c = 0.189$ GeV for 2 flavor, $T=T_c = 0.128$ GeV for 2+1 flavor ,and $T=T_c = 0.131$ GeV for 2+1 flavor. From the plots, it can be observed that the running coupling constant for each system initially exhibits an upward trend with $L$, culminating in a maximum value before gradually decreasing. Furthermore, as the chemical potential increases, this maximum value progressively diminishes. These plots provide further evidence that with increasing temperature and chemical potential, heavy quarkonium becomes more unstable.

\begin{figure}
    \centering
    \includegraphics[width=8.5cm]{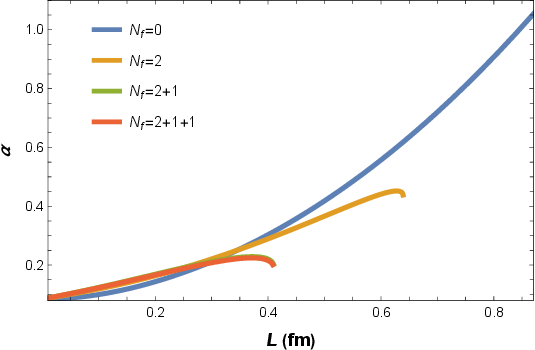}
    \caption{\label{fig17} The running coupling constant is dependent on quark separation distance when $T=0.2$ GeV, $\mu=0$.}
\end{figure}

\begin{figure}
    \centering
    \includegraphics[width=8.5cm]{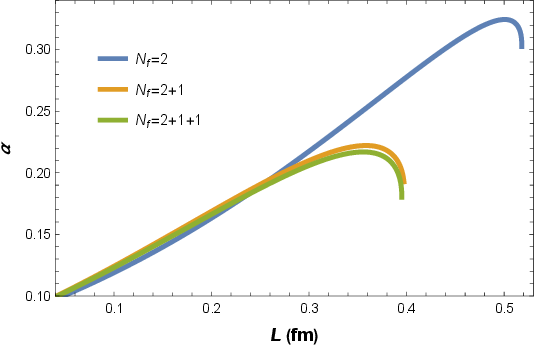}
    \caption{\label{fig18} The running coupling constant is dependent on quark separation distance when $T=0.2$ GeV, $\mu=0.3$ GeV.}
\end{figure}

Furthermore, to explore the running coupling constants of different flavors, we have generated coupling constant plots for various systems under the same temperature and chemical potential, as illustrated in Figs.~\ref{fig17} and~\ref{fig18}. A discernible trend emerges: as the separation distance between quarks increases, the coupling constant initially rises before decreasing. Additionally, with an increase in flavors, we observe a reduction in the maximum value of the coupling constant. In Fig.~\ref{fig17}, corresponding to increasing flavors, the coupling constant values are 0.452, 0.224, and 0.227, respectively. Similarly, in Fig.~\ref{fig18}, the coupling constant values are 0.326, 0.222, and 0.217 following the same pattern. These results indicate that as the system's flavor grows, the heavy quarkonium becomes more unstable, leading to an earlier manifestation of unbound quark pairs.


\section{The real-time dynamics of quark dissociation for different flavors}
\label{sec5}

We now turn our attention to the thermalization process of quark-antiquark pairs. At the boundary, a pair of heavy quark and antiquark is generated with a fixed distance $L$ between them. If the quarks are sufficiently separated at a given temperature, the gluon cloud surrounding the quark pair will gradually thermalize, merging with the surrounding medium. Consequently, the heavy quark state will undergo dissociation. The holographic description of this dissociated quark pair involves a string with fixed endpoints at the boundary and extending towards the horizon of a black hole. As the string falls under the gravitational influence of the background, it eventually reaches the black hole horizon and attains equilibrium, transforming into a straight string plummeting into the black hole \cite{Iatrakis:2015sua,Jena:2022nzw,Bellantuono:2017msk}.

In this section, we delve into the real-time dynamics of quark pairs with different flavors. The string's coordinates are represented as $X^M=\left(t, z\left(t, x\right), x, 0,0\right)$. The Nambu-Goto action then reads
\begin{eqnarray}
S_{N G}=\frac{1}{2 \pi T} \int d t d x \frac{e^{2 A_s(z)}}{z^2} \sqrt{g(z)+z^{\prime 2}-\frac{\dot{z}^2}{g(z)}},
\end{eqnarray}
where $\dot{z}=\partial_t z $ and $z^{\prime}=\partial_{x} z$. Then, the motion equation of the string can be written as
\begin{eqnarray}
\begin{gathered}
-\partial_t\left(\frac{e^{2 A_s(z)}}{z^2 g(z)} \frac{\dot{z}}{\sqrt{g(z)+z^{\prime 2}-\frac{\dot{z}^2}{g(z)}}}\right)+\partial_{x}\left(\frac{e^{2 A_s(z)}}{z^2} \frac{z^{\prime}}{\sqrt{g(z)+z^{\prime 2}-\frac{\dot{z}^2}{g(z)}}}\right) \\
-\partial_z\left(\frac{e^{2 A_s(z)}}{z^2}\right) \sqrt{g(z)+z^{\prime 2}-\frac{\dot{z}^2}{g(z)}}-\frac{e^{2 A_s(z)} \partial_z g(z)}{2 z^2 \sqrt{g(z)+z^{\prime 2}-\frac{\dot{z}^2}{g(z)}}}\left(1+\frac{\dot{z}^2}{g(z)^2}\right)=0.
\label{eq:motion}
\end{gathered}
\end{eqnarray}
We can numerically solve the above equation with suitable boundary conditions. To achieve this, we fix the endpoints of the string on the boundary at point $z\left(t, x= \pm L / 2\right)=\varepsilon_{U V}$. Additionally, we assume that the string is initially at rest on the boundary at point $z\left(t=0, x\right)=\varepsilon_{U V}$ with a velocity of $\dot{z}\left(t=0, x\right)=\varepsilon_{U V}$. Here, $\varepsilon_{U V}$ represents the boundary cutoff, and we set $\varepsilon_{U V} = 0.11$ in our case. In principle, using these boundary conditions, we can solve Eq.~\eqref{eq:motion} for all values of $L$. However, for small separation of quark pairs, the spatial variation component introduces significant numerical errors, making it challenging to solve this partial differential equation. The difficulty is further amplified by the singularity of our coordinate system at the black hole horizon, rendering it unsuitable for studying dynamics near the horizon. Therefore, meticulous numerical care is required when discussing dynamics for small $L$. On the other hand, for large quark separations $L$, the string profile approximates a straight line falling towards the black hole horizon. In this case, the spatial variation becomes negligible, meaning that the dependence on $x$ can be effectively ignored for most of the string. This bypasses the aforementioned numerical intricacies. This simplification greatly facilitates the understanding of real-time dynamics of the string. Furthermore, when considering quark dissociation (primarily described by the disconnected string configuration in the holographic context), it is indeed reasonable to focus on dynamics governed by large $L$ physics \cite{Jena:2022nzw}.

\begin{figure}
    \centering
    \includegraphics[width=8.5cm]{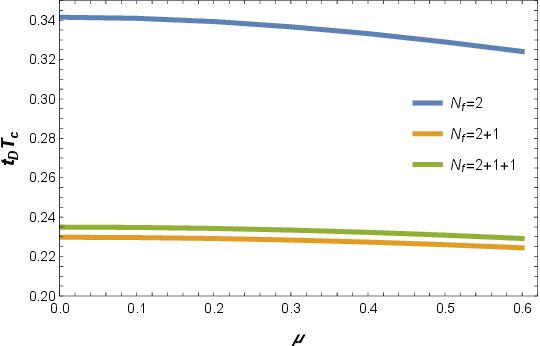}
    \caption{\label{fig19} The quark dissociation time is a function of chemical potential for quarkonium pairs of different flavors when $T=0.3$ GeV. $T_{c}$ is the critical temperature, which can be found in Table~\ref{table:table1}.}
\end{figure}

\begin{figure}
    \centering
    \includegraphics[width=8.5cm]{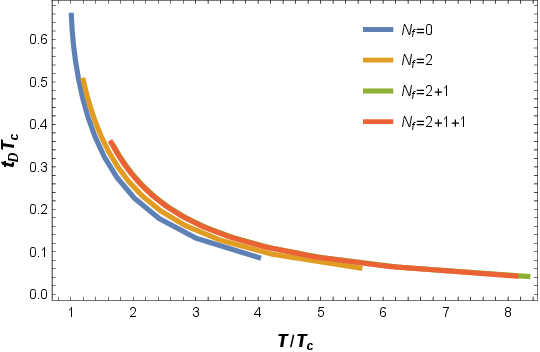}
    \caption{\label{fig20} The quark dissociation time is a function of temperature for quarkonium pairs of different flavors when $\mu=0$. $T_{c}$ is the critical temperature, which can be found in Table~\ref{table:table1}.}
\end{figure}

In this simpler case, the string action takes the form $S_{N G}=\frac{1}{2 \pi } \int d t d x \frac{e^{2 A_s(z)}}{z^2} \sqrt{g(z)-\frac{\dot{z}^2}{g(z)}}$. Then, the equation of motion for the string becomes,
\begin{eqnarray}
\begin{aligned}
& -\partial_t\left(\frac{e^{2 A_s(z)}}{z^2 g(z)} \frac{\dot{z}}{\sqrt{g(z)-\frac{\dot{z}^2}{g(z)}}}\right)-\partial_z\left(\frac{e^{2 A_s(z)}}{z^2}\right) \sqrt{g(z)-\frac{\dot{z}^2}{g(z)}} \\
& -\frac{e^{2 A_s(z)} \partial_z g(z)}{2 z^2 \sqrt{g(z)-\frac{\dot{z}^2}{g(z)}}}\left(1+\frac{\dot{z}^2}{g(z)^2}\right)=0.
\end{aligned}
\end{eqnarray}
The action does not depend on time, therefore there exists a corresponding conserved energy. The Lagrange can be written as
\begin{eqnarray}
\mathcal{L} =\frac{1}{2 \pi }  \frac{e^{2 A_s(z)}}{z^2} \sqrt{g(z)-\frac{\dot{z}^2}{g(z)}}.
\end{eqnarray}
Because of the definition of Hamiltonian as
\begin{eqnarray}
\mathcal{H} = \dot{z} \frac{\partial{\mathcal{L}}}{\partial{\dot{z}}} - \mathcal{L},
\end{eqnarray}
the conserved energy $E$ can be derived:
\begin{eqnarray}
\mathcal{H} = T_s \frac{e^{2 A_s(z)}}{z^2} \frac{g(z)}{\sqrt{g(z)-\frac{\dot{z}^2}{g(z)}}}=E.
\end{eqnarray}
Here, $T_s=1 / 2 \pi$ is the open string tension. Therefore, the velocity of the string can be expressed as
\begin{eqnarray}
\dot{z}=\frac{g(z)}{E} \sqrt{\left(E\right)^2-\frac{T_s^2 e^{4 A_s(z)} g(z)}{z^4}}.
\end{eqnarray}
We can determine $E$, the energy of the falling string, from the boundary conditions of the string being dropped with zero initial velocity ($\dot{z}(t=0)=0$). By setting the boundary cutoff value at $z=\varepsilon_{U V}$, the energy of the falling string can be represented as
\begin{eqnarray}
E=\frac{T_s e^{2 A_s\left(\varepsilon_{U V}\right)} \sqrt{g\left(\varepsilon_{U V}\right)}}{\varepsilon_{U V}^2}.
\end{eqnarray}
The time required for the dissociation of a quarkonium can be calculated as the time it takes for the string to approach the event horizon from the boundary. As the string asymptotically approaches the horizon, this time becomes infinite. To avoid this divergence, we introduce an infrared cutoff distance $\varepsilon_{I R}=0.01$ from the horizon, which we consider as the point of string thermalization. Therefore, we solve the string motion equation corresponding to different black hole backgrounds at various temperature and chemical potential states and calculate the dissociation time $t_D$. It is straightforward to obtain a closed-form expression for $t_D$ from the string's motion equation
\begin{eqnarray}
t_D=\int_{\varepsilon_{U V}}^{z_h - \varepsilon_{I R}} d z \frac{1}{g(z) \sqrt{1-\frac{T_s^2 e^{4 A_s(z)} g(z)}{(E)^2 z^4}}}.
\end{eqnarray}
Based on our investigation into the quark pair dissociation time as a function of $\mu$ at various flavors, we have generated Fig.~\ref{fig19} to visualize the findings. It is evident from the graph that the dissociation time of quark pairs gradually decreases with an increase in chemical potential for different flavors. Moreover, the quark pair dissociation time is found to be the longest for $N_{f}=2$ and the shortest for $N_{f}=2+1+1$, implying that an increase in flavor leads to a reduction in the dissociation time. In Fig.~\ref{fig20}, we present the dissociation time of different flavor quark pairs at varying temperatures. As the temperature rises, the dissociation time for each quark pair experiences a significant decrease. This phenomenon aligns with our expectations from physics, as larger black holes cause strings to descend more rapidly towards the horizon, resulting in accelerated string dissociation. Similarly, an increase in flavor also contributes to a reduction in the dissociation time. However, compared to Fig.~\ref{fig19}, the introduction of temperature intensifies the decrease in dissociation time, signifying that temperature holds a more substantial influence than the chemical potential.

\section{Conclusion}
\label{sec6}

The main focus of this article is to investigate the potential energy, dissociation distance, running coupling, and dissociation time of quark-antiquark pairs in different systems, namely the pure gluon system, 2-flavor system, 2+1-flavor system, and 2+1+1-flavor system at finite temperature and chemical potential with a holographic model. Our results reveal that both high temperature and chemical potential cause a decrease in the maximum separation distance of quark-antiquark pairs. The strings connecting the quark pairs break at smaller separation distances, leading to the transition of quark-antiquark pairs into the free quarks. Under the same temperature and chemical potential, an increase in the flavor of quark pairs reduces the dissociation distance. Additionally, we observed that the inclusion of chemical potential and temperature leads to a reduction in the linear potential energy component of quark-antiquark pairs, while the Coulombic potential energy component remains unaffected.

Furthermore, we have conducted a study on the dissociation distance and running coupling constant of quark-antiquark pairs under temperature and chemical potential conditions. Our findings reveal that temperature has a more significant influence on quark-antiquark pairs compared to chemical potential of the same magnitude. Furthermore, our computational results provide further evidence that the inclusion of temperature and chemical potential renders quark pairs more unstable, leading to their dissociation occurring at shorter quark separation distances.

We have also investigated the real-time dynamics of quark pair dissociation, where the dissociation scenario corresponds to strings falling from the boundary to the horizon. We have computed the dissociation time for heavy quark separation at various temperatures and chemical potentials and observed that the dissociation time decreases as the temperature and chemical potential increase. Besides, we have specifically examined the dissociation time of quark pairs in different flavor systems under the influence of high temperature and chemical potential. The outcomes have demonstrated that the dissociation time of heavy quarkonium increases with an increase in the number of flavors.

\section*{ACKNOWLEDGMENTS}
This work is supported by the Natural Science Foundation of Hunan Province of China under Grant No. 2022JJ40344, the Research Foundation of Education Bureau of Hunan Province, China (Grant No. 21B0402), and the National Natural Science Foundation of China Grants No. 12175100 and No.12350410371.

\section*{References}
\bibliography{ref}

\end{document}